\documentclass[twocolumn,pra,superscriptaddress,amssymb,amsmath,amsmath,floatfix]{revtex4-1}
\usepackage{graphicx}
\usepackage{epstopdf}
\usepackage{braket}
\usepackage{bm}
\usepackage[colorlinks=true,linkcolor=blue,urlcolor=blue,citecolor=blue]{hyperref}
\usepackage{comment}
\usepackage{float}
\usepackage{cancel}
\usepackage{times}
\usepackage{amsmath, amsthm, amssymb, amsfonts}
\usepackage{xcolor}
\usepackage{subcaption}
\newcommand{\affFUW}{Faculty of Physics, University of Warsaw, Pasteura 5, 02-093 Warsaw, Poland}

\newcommand{\beq}{\begin{equation}}
	\newcommand{\eeq}{\end{equation}}

\newcommand{\oper}[1]{\mathbf{#1}} 
\newcommand{\opp}[1]{#1} 

\begin{document}
	\author{Sakthikumaran Ravichandran}
	\affiliation{\affFUW}
	\author{Piotr Kulik}
	\affiliation{\affFUW}
	\author{Krzysztof Jachymski}
	\affiliation{\affFUW}
	\date{\today}
	
	\title{Quantum engineering with ultracold polar molecules using trap-induced resonances}
	
	\begin{abstract}
		Polar molecules represent a promising platform for quantum simulation and computation protocols. Highly controllable arrays of optical tweezers are now accessible in experiments, allowing for unprecedented control of individual molecules. Motional dephasing is typically seen as an obstacle in quantum computing scenarios. Here, we instead consider using the trap structure as a resource for implementing efficient quantum gates. By numerically solving the two-body problem of dipoles trapped in separate tweezers, we identify trap-induced resonances that can serve as the mechanism for achieving state-dependent dynamics and can be further utilized for quantum sensing.
	\end{abstract}
	
	\maketitle

	\section{Introduction}

	Ultracold polar molecules have rapidly advanced to become a leading platform for quantum science, offering long-lived internal states, tunable dipole-dipole interactions, and compatibility with microscopic trapping architectures. These features have driven progress in quantum simulation, quantum computation, precision measurements and controlled chemistry~\cite{Blackmore2018, Bohn2017, Safronova2018, Liu2021, Dulieu2017, Softley2023,Cornish2024}. Unlike atoms, polar molecules possess a permanent electric dipole moment, enabling strong, anisotropic interactions whose magnitude and sign can be engineered using static or microwave fields. This tunability provides direct access to extended-range spin Hamiltonians, programmable interactions, and coherent dipolar exchange, key ingredients for constructing scalable quantum devices. Microwave control has also enabled the realization of molecular Bose-Einstein condensate by suppressing inelastic collisions~\cite{Bigagli2024}.
	
	A significant milestone in this field has been the development of optical tweezer arrays for molecules, which combine single particle control with dynamically reconfigurable geometries. Recent experiments have demonstrated tweezer trapping and laser cooling of CaF~\cite{Anderegg2019, Gregory2021}, as well as assembly of single ground-state NaCs and RbCs molecules from individually trapped atoms~\cite{Cairncross2021, Ruttley2024}. Complementary work has shown site-resolved state readout, microwave-driven multilevel manipulation, and defect-free array preparation with full internal quantum state control~\cite{Picard2024}. Crucially, Raman sideband cooling techniques have been adapted to polar molecules, allowing for near-ground state cooling~\cite{Bao2024,Lu2024}. These achievements position molecular tweezer arrays alongside Rydberg atoms as a flexible quantum platform. Beyond state preparation and control, these capabilities are now translating into tweezer-based molecular quantum simulators. Since the seminal achievement of direct detection of dipolar spin-exchange interactions in optical lattice~\cite{Yan2013}, there has been substantial technological progress, increasing the number of particles and lattice filling, and adding more control tools. Coherent dipolar spin exchange has been used to realize an effective spin-$1/2$ XY Hamiltonians and to generate entanglement between molecules~\cite{Bao2023, Holland2023}, directly illustrating the route to programmable spin model dynamics and beyond, reaching further to generalized t-J model~\cite{Carroll2025} and Floquet engineering~\cite{Miller2024,Cheuk2026}.

	Strong, anisotropic dipole-dipole interactions of polar molecules are a natural resource for fast entangling gates. DeMille's pioneering work~\cite{DeMille2002} proposed encoding qubits in molecular rotational states in a 1D array and utilizing electric-field-tunable dipolar interactions to mediate two-qubit logic  gates. Subsequent theoretical studies extended this concept to other lattice configurations, demonstrating that microwave dressing and state-dependent dipole moments facilitate robust entangling operations between molecular qubits~\cite{Micheli2006, Buchler2007, Yelin2006,Ni2018}. However, realizing fully programmable interactions requires understanding how dipolar couplings, external fields, and trapping potentials combine to reshape the two-body spectrum under realistic conditions. In particular, fluctuations in tweezer positions, motional dephasing, and state-dependent light shifts can limit gate fidelities in tightly confining traps~\cite{Hughes2019,Bergonzoni2025}. Dependence of the trap frequency on the internal state results in couplings that complicate the structure of the problem, but can be mitigated using so-called magic wavelengths for which the polarizability is the same. Combined with long lifetimes of rotational states, this enables second scale coherence and long-lived entanglement, leading to deeper attainable digital circuits~\cite{Ruttley2024, Picard2024, Ruttley2025}. 
	
	The structure of trapped interacting ultracold atoms is well understood in general, from the most basic problem of short-ranged interaction~\cite{Busch1998}, to including separate traps~\cite{Stock2006,Krych2009}, and finite range interactions~\cite{Kanjilal2007,Grishkevich2009,Schulz2015,wall2013dipole} with a multichannel generalization~\cite{Dawid2018}. Trap-induced resonances (TIRs) can emerge here as avoided crossings between vibrational states dominated by the harmonic trap and short range molecular bound states. They are crucial for mergoassociation of molecules from separate atoms~\cite{Jachymski2013,Bird2023}, and offer a robust mechanism for amplifying and controlling the interactions~\cite{Stock2006}. TIRs are a general phenomenon which does not require specific type of interactions or trap structure and are always expected when separation between two confined particles is varied, but their detailed properties strongly depend on the specific system.
	
\begin{figure}
		\centering
		\includegraphics[width=\linewidth]{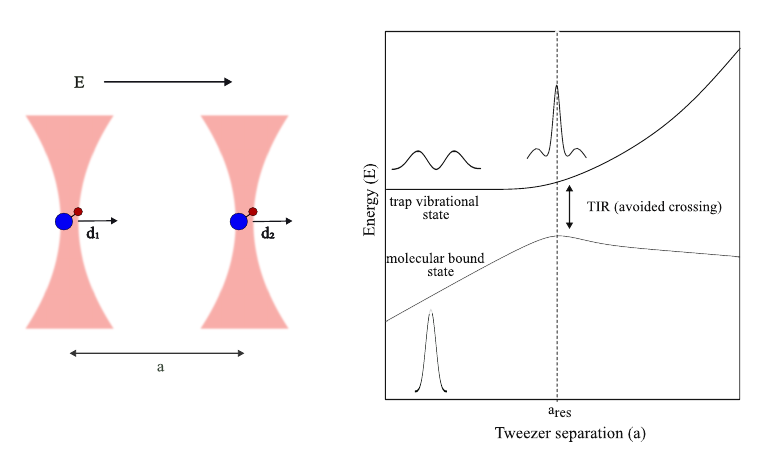}
		\caption{ Left: the setup consisting of two molecules trapped in individual optical tweezers, with an electric field pointing along the intermolecular axis. Right: the basic mechanism behind the occurrence of a trap-induced resonance.
		}
		\label{fig:basics}
    	\end{figure}

	In this work, we investigate state-dependent trap-induced resonances between ultracold polar molecules in optical tweezers, visualized in Figure~\ref{fig:basics}. This problem has been analyzed in a simplified setting by~\cite{Sroczynska2021}, assuming quasi-1D geometry and employing exact diagonalization in the noninteracting single-particle basis, which makes the identification of resonances numerically inefficient. Here, we focus on experimentally realistic parameters and interaction potentials, and turn to a coupled-channel treatment based on the renormalized Numerov method~\cite{Johnson1978, DU1993}. We resolve the two-molecule energy structure across a wide range of trap separations and electric field strengths, showing that TIRs appear naturally for experimentally relevant parameters and give rise to strong, controllable interaction shifts. We demonstrate that the internal state dependence of dipole-dipole couplings leads to rotational state selective resonances. Building on these results, we derive an effective Hamiltonian that maps the full molecular interaction landscape onto a reduced two qubit basis, enabling a compelling design of dipolar entangling gates. We further outline how these resonances may be used for the purpose of electric field sensing.
	
	The remainder of the paper is structured as follows. Section II introduces the molecular Hamiltonian and the qubit space, and presents numerical diagonalizations and characterizations of trap-induced resonances. Section III details the numerical results, including trap-induced resonances, the gate protocol, and sensing applications. Section IV concludes with an outlook.

	\section{System of interest}
	
	\subsection{Hamiltonian and qubit space}
	
	In the system we consider, each molecule is modelled as a rigid rotor, which is a well-established approximation in rotational spectroscopy~\cite{Brown}. We disregard the hyperfine structure details, as they do not fundamentally change the physics that we want to focus on and they are straightforward to include if needed for realistic implementation~\cite{Blackmore2023,Humphreys2025}, providing further opportunities for qubit or qudit encoding. The internal rotational dynamics is captured by the Hamiltonian
	
	\beq\label{Hrot}
	H_{\rm rot}=B \oper{j}^{2}-\opp{d}_0 E,
	\eeq
	
	where $B$ is the rotational constant, $\mathbf{j}$ is the rotational angular momentum operator, and $d_0 = \hat{e}_0 \cdot \mathbf{d}$ is the projection of the molecule's electric dipole moment onto the quantization axis set by an external static electric field $\mathbf{E}$. This Hamiltonian describes the relationship between free rotational motion and orientation caused by the field. 
	The eigenstates of the field-free rigid rotor are the spherical harmonics$\ket{j, m}$, with energy eigenvalues $B j(j+1)$. When an electric field is applied, these states are no longer eigenstates of the complete Hamiltonian due to the term  $-d_0 E$, which mixes states with different $j$ but the same $m$. This leads to Stark mixing and creates pendular states that can be controlled by adjusting the field strength. To measure this interaction, we define a dimensionless parameter $\beta = dE/B$, which reflects the Stark interaction strength compared to the rotational splitting.

	Considering a pair of molecules, the two-body  potential is well approximated by the electric dipole-dipole interaction. It stems from the coupling between two molecular dipole moments and is conveniently represented in spherical tensor form as
	
	\beq
	V_{\rm dd}=-\frac{\sqrt{6}}{r^{3}}\sum_{p=-2}^{2}{(-1)^{p}C_{-p}^{2}(\theta,\phi)T_p ^{2}(\oper{d}_1,\oper{d}_2)}\, 
	\eeq
	
	where $C^2_p(\theta, \phi)$ are Racah-normalized spherical harmonics and $T^2_p$ are rank-2 tensor operators acting on the dipoles. For aligned molecules, where the quantization axis aligns with the intermolecular axis, the primary energy-conserving term is  $p = 0$, resulting in a simplified interaction
	
	{\small
		\beq
		\begin{split}
			V_{\rm dd}^{p=0}
			&=
			-\frac{1}{r^{3}}
			\bigl(3\cos^{2}\theta-1\bigr)
			\Bigl[
			\opp{d}_0^{(1)}\opp{d}_{0}^{(2)}
			+\frac{1}{2}\opp{d}_+^{(1)}\opp{d}_{-}^{(2)}
			+\frac{1}{2}\opp{d}_-^{(1)}\opp{d}_{+}^{(2)}
			\Bigr]
		\end{split}
		\label{interaction}
		\eeq
	}
	
	This coupling is fundamental for practically all two-qubit gates discussed in the literature. It features long-range nature due to slow power-law decay and angular dependence, which can be used for additional control. Note that the total angular momentum projection is conserved, and the $d_+ d_-$ term naturally realizes an exchange interaction, which can lead to a SWAP gate~\cite{Ni2018}. In this work, we are rather interested in implementing a controlled phase gate, so we will be looking at state-dependent dynamics.
	
	To define a qubit basis, we choose two well isolated rotational states. It is convenient to work with the lowest  lying manifold within rotational-hyperfine sublevels, which exhibit very long coherence times. Using magic wavelengths it is possible to reduce the motional decoherence~\cite{Gregory2021} while maximizing dipolar coupling~\cite{Bao2023}. We treat these internal states as pseudospins, forming the qubit computational basis $\{\ket{0}, \ket{1}\}$.
	
	\subsection{Mapping onto the two-qubit manifold}
	
	We now detail how the microscopic two-molecule Hamiltonian maps onto a qubit description. We define a single-molecule qubit by selecting two long-lived internal states $\{\ket{0},\ket{1}\}$, typically chosen among field-dressed rotational eigenstates of $\hat H_{\rm rot}$ \cite{Micheli2006,DeMille2002,Gorshkov2011b}. The corresponding two-qubit computational basis is
	\begin{equation}
		\mathcal{P}\equiv \mathrm{span}\{\ket{00},\ket{01},\ket{10},\ket{11}\},
	\end{equation}
	Within this framework, the dipole-dipole interaction $\hat V_{\rm dd}$ generates both (i) diagonal interaction shifts that depend on the induced dipoles of the selected qubit states and (ii) off-diagonal exchange couplings whenever the transition dipole between $\ket{0}$ and $\ket{1}$ is allowed \cite{Gorshkov2011a,Gorshkov2011b,Hazzard2014}. Specifically, in the aligned quantization geometry described above, the $d^{(1)}_{\pm}d^{(2)}_{\mp}$ terms mediate processes of the form $\ket{01}\leftrightarrow\ket{10}$ (dipolar flip flop), while the $d^{(1)}_{0}d^{(2)}_{0}$ term produces state-dependent energy shifts \cite{Gorshkov2011b,Lahaye2009,Krems2008}. The resulting interaction structure is formally equivalent to that of effective spin models implemented with polar-molecule pseudospins, in which diagonal (Ising-like) and off-diagonal (flip flop) terms originate from different tensor components of the dipolar interaction \cite{Micheli2006,Buchler2007,Gorshkov2011b}.

	While a dc electric field is commonly introduced as a means to polarize molecules and increase the interaction strength by inducing a finite dipole moment~\cite{DeMille2002,Ospelkaus2010}, it simultaneously reshapes the internal spectrum by Stark mixing and can therefore increase the number and relevance of intermediate channels coupled by $\hat V_{\rm dd}$~\cite{Wall2015,Bohn2017}. This effect is especially significant for qubit encodings based on rotational states. The dc field modifies both permanent dipoles $\langle 0|d_0|0\rangle$, $\langle 1|d_0|1\rangle$ and transition dipoles $\langle 0|d_q|1\rangle$. It also changes the detunings to other rotational manifolds, which give rise to effective interactions at second and higher orders. Therefore, increasing $E$ does not always result in a steady improvement of the gate Hamiltonian. While higher fields can enhance desired couplings, they may also increase susceptibility to leakage and render the interaction landscape more multichannel and more sensitive to parameter drifts, such as field noise, polarization imperfections, and state-dependent light shifts~\cite{Gorshkov2011b,Wall2015}. This trade-off between stronger interactions and increased state mixing accounts for the fact that the dc field serves as both a constraint and a control parameter in practical tweezer-based experiments.

	We use Schrieffer-Wolff (SW) transformation \cite{Schrieffer1966} to describe the qubit space while still accounting for virtual transitions into higher rotational and motional channels. The full Hilbert space is divided into the computational manifold $P$ (spanned by $\ket{00},\ket{01},\ket{10},\ket{11}$) and its complement $Q=\mathbf{1}-P$ \cite{Bravyi2011a}. We briefly summarize the SW construction used to define the effective qubit-qubit interaction, establish notation, and identify the virtual channels contributing at second order. We decompose the dipolar interaction $\hat V$ into diagonal and off-diagonal parts with respect to $(P,Q)$,	
    \begin{equation}
    	\begin{gathered}
	    	\hat V = \hat V_{\rm diag} + \hat V_{\rm od} \\
		    \begin{aligned}
		    	\hat V_{\rm diag} &\equiv P\hat V P + Q\hat V Q
		    	\qquad
		    	\hat V_{\rm od} &\equiv P\hat V Q + Q\hat V P
	    	\end{aligned}
    	\end{gathered}
	    \label{eq:V_diag_od}
    \end{equation}
	
	The SW transformation is implemented by a unitary rotation generated by an anti-Hermitian operator $\hat S=-\hat S^\dagger$,
	\begin{equation}
		\hat H_{\rm SW}=e^{\hat S}\hat H e^{-\hat S}\, ,
	\end{equation}
	chosen such that $\hat H_{\rm SW}$ becomes block-diagonal at each order in $\hat V_{\rm od}$. At the first order, this means the transformed Hamiltonian should not have any $P$-$Q$ couplings, which gives the defining equation for the generator,
	\begin{equation}
		[\hat H_0+\hat V_{\rm diag},\,\hat S] = -\,\hat V_{\rm od}\, .
		\label{eq:SW_generator_Vdiag}
	\end{equation}
	Expanding $\hat H_{\rm SW}$ to second order in $\hat V_{\rm od}$ gives the effective Hamiltonian within the qubit manifold \cite{Bravyi2011a},
	\begin{equation}
		\hat H_{\rm eff}
		= P(\hat H_0+\hat V_{\rm diag})P
		+ \frac{1}{2}\,P[\hat S,\hat V_{\rm od}]P
		+ \mathcal{O}(\hat V_{\rm od}^3)
		\label{eq:Heff_SW_commutator}
	\end{equation}
	For practical evaluation it is often useful to write the second-order term in matrix-element form. For qubit states $\ket{\alpha},\ket{\beta}\in P$ and intermediate states $\ket{k}\in Q$ (eigenstates of $\hat H_0+\hat V_{\rm diag}$ with energies $E_k$), one 
	\begin{equation}
		\begin{split}
			\bigl(H_{\rm eff}\bigr)_{\alpha\beta}
			&=
			(E_\alpha)\delta_{\alpha\beta}
			+
			(V_{\rm diag})_{\alpha\beta} \\
			&\quad+
			\frac{1}{2}\sum_{k\in Q}
			V_{\alpha k}V_{k\beta}
			\left(
			\frac{1}{E_\alpha-E_k}
			+
			\frac{1}{E_\beta-E_k}
			\right),
		\end{split}
		\label{eq:Heff_matrix_elements_sym}
	\end{equation}
	
	where $E_\alpha$ denotes the unperturbed qubit-manifold energy from $\hat H_0+\hat V_{\rm diag}$ and
	$V_{\alpha k}\equiv \bra{\alpha}\hat V_{\rm od}\ket{k}$. In particular, the diagonal shift of a computational basis state $\ket{\alpha}$ reduces to \cite{Cohen1998,Krems2005}

	\begin{equation}
		\Delta E_\alpha^{(2)}
		=
		-\,\sum_{k\in Q}
		\frac{\bigl|\bra{\alpha}\hat V_{\rm od}\ket{k}\bigr|^2}{E_k-E_\alpha}\, .
		\label{eq:Heff_diag_shift}
	\end{equation}

	Equation~\eqref{eq:Heff_diag_shift} gives the general second-order energy shift induced by virtual dipolar couplings. In the far detuned regime, where all intermediate channels are nonresonant, the shift takes the van der Waals form	$\Delta E_\alpha^{(2)}(R) = -C_6^{(\alpha)}/R^6$ with the state-dependent coefficient~\cite{Karman2024}
	
	\begin{equation}
		C_6^{(\alpha)}
		=
		\sum_{k\in Q}
		\frac{\bigl|\bra{\alpha}\hat V_{\rm od}(R^3)\ket{k}\bigr|^2}{E_k-E_\alpha},
		\label{eq:C6_def}
	\end{equation}
	
	where the explicit $R^{-3}$ scaling of $\hat V_{\rm dd}$ has been factored out. The coefficient $C_6^{(\alpha)}$ therefore results from contracting two dipole-dipole couplings through the manifold of intermediate pair states.
	
	The dominant contribution to $C_6$ often arises from virtual rotational excitations within the electronic ground state \cite{Karman2024}. Because rotational splittings are small compared to electronic excitation energies, this rotational contribution can exceed the electronic dispersion component. Two molecules prepared in identical rotational states typically exhibit attractive van der Waals interactions, whereas molecules in sufficiently different rotational states can also interact repulsively. When near-degenerate exchange channels are present, such as rotational quantum numbers differing by one, the interaction crosses over to resonant dipole exchange with leading $R^{-3}$ scaling, beyond the purely perturbative $R^{-6}$ description.
	 
	In practice, different computational basis states acquire different van der Waals coefficients $C_6^{(\alpha)}$, leading to state-dependent diagonal shifts. In addition, the dipolar tensor structure of $\hat V_{\rm dd}$ also produces off-diagonal exchange terms of the form $\ket{01}\!\bra{10}+\ket{10}\!\bra{01}$. The effective two-qubit Hamiltonian contains both density-density (Ising-like) and flip-flop contributions, whose relative strength is set by the rotational structure and the applied dc field.
	
\begin{figure}[H]
	    	\centering
	    	
	    	\begin{subfigure}{0.85\linewidth}
	    		\centering
	    		\includegraphics[width=\linewidth]{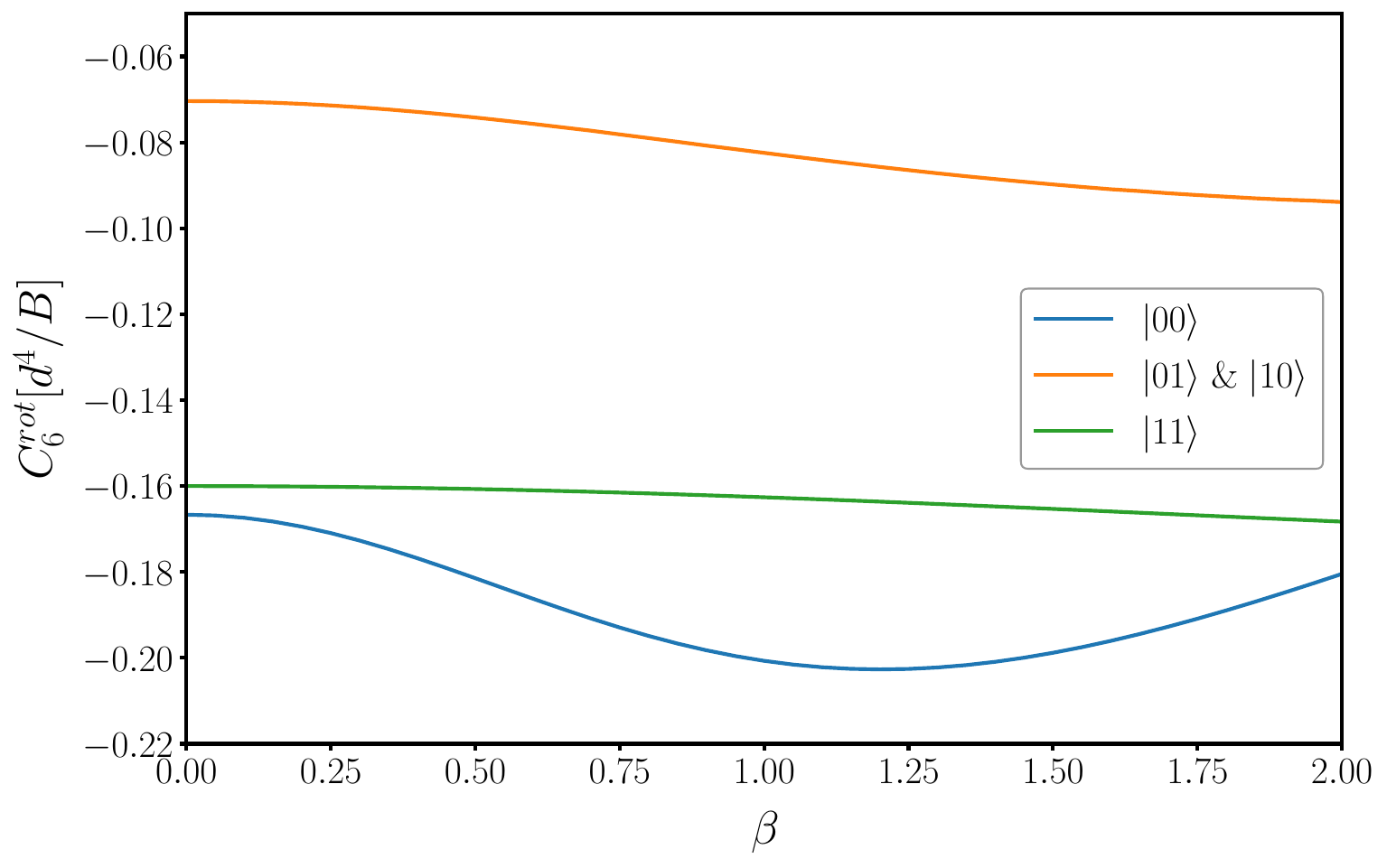}
	    		\caption{$C_6$ coefficients for all possible qubit pair states as a function of dimensionless electric field strength.}
	    	\end{subfigure}
	    	
	    	\vspace{0.5cm}
	    	
	    	\begin{subfigure}{0.85\linewidth}
	    		\centering
	    		\includegraphics[width=\linewidth]{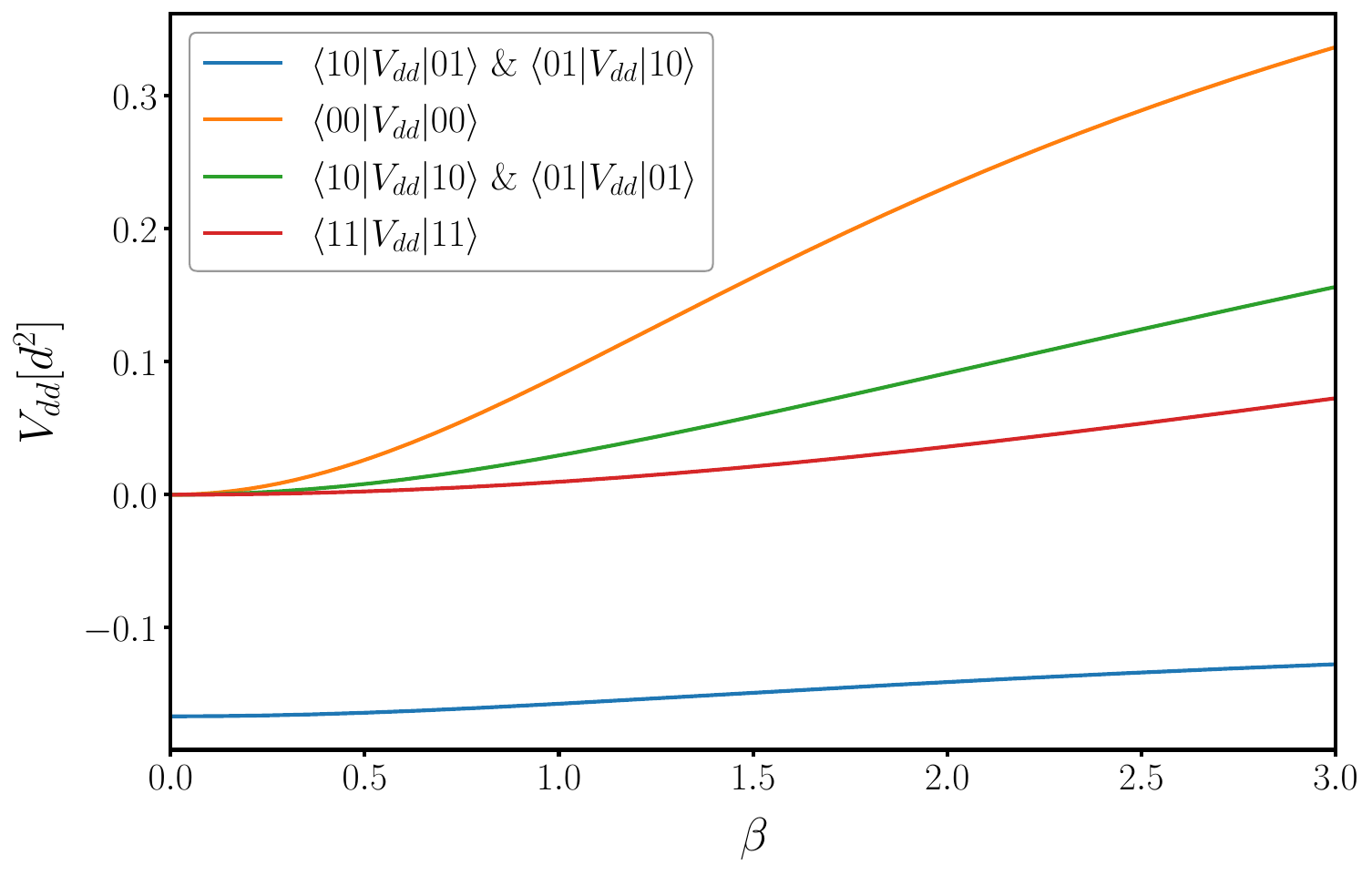}
	    		\caption{Effective dipolar interaction for the same case.}
	    	\end{subfigure}
	    	
	    	\caption{Interaction properties for the considered qubit pair states, expressed in dimensionless form.}
	    	\label{fig:int}
	    \end{figure}

	Figure~\ref{fig:int}a shows the strength of the van der Waals interaction as a function of the applied electric field in natural units of $d^4/B$. For a realistic system involving e.g. NaCs molecules this corresponds to $C_6$ values in the range of $10^7$ atomic units. The resulting characteristic van der Waals length $R_6=(2\mu C_6/\hbar^2)^{1/4}\approx2\cdot 10^3$ Bohr radii. Thus, even without the presence of electric field inducing a dipole moment, the molecules may already be strongly coupled. To characterize the strength of dipolar interactions, one can introduce another lengthscale $R_3=2\mu \langle d^2\rangle /\hbar^2$, where $\langle d^2\rangle$ is the induced dipole moment for a particular pair of states in the presence of electric field. In Fig.~\ref{fig:int}b, we showcase field tuning of dipolar interactions for various pair states.

    \subsection{Including the tweezers}\label{sec:incl_tweezers}
    
    The molecules are trapped in optical tweezers, so we also need to include their motion. We approximate the traps as separate harmonic potentials, which should apply for sufficiently low temperatures.  Recent experimental demonstration of Raman sideband cooling~\cite{Bao2024,Lu2024} validates the assumption that the molecular pair can be prepared close to the motional ground state. The full Hamiltonian of two interacting molecules in tweezers takes the form
    \begin{equation}
        \begin{split}
            \hat{H} =& -\frac{\hbar^2}{2m_1}\Delta_{r_1} + \frac{1}{2}m_1\omega^2((z_1 - a_1)^2+\eta\rho_1^2) \\
        	&- \frac{\hbar^2}{2m_2}\Delta_{r_2} + \frac{1}{2}m_2\omega^2((z_2-a_2)^2 + \eta^2\rho_2^2)\\
            &- \frac{C_6}{|\Vec{r_2} - \Vec{r_1}|^6} 
            - \frac{3\Vec{d_1}\cdot(\Vec{r_2} - \Vec{r_1})\Vec{d_2}\cdot(\Vec{r_2} - \Vec{r_1}) - \Vec{d_1}\Vec{d_2}}{4\pi\varepsilon_0 |\Vec{r_2} - \Vec{r_1}|^3}. \\
        \end{split}
    \end{equation}
    Here $\omega$ is the trapping frequency along the $z$ axis, which is set by the orientation of the dipoles and for simplicity we assume the trap displacement to be parallel to $z$ in order to preserve cylindrical symmetry. This ensures conservation of the orbital angular momentum projection $m_z$, reducing the size of the problem. Furthermore, $\eta$ is the trap anisotropy parameter, and $a_1$ and $a_2$ are the tweezer positions. After center of mass transformation, the system becomes separated and we can focus only on the physically significant relative part that only depends on the distance between the tweezers $a=a_2-a_1$. 
    \begin{equation}\label{eq:H_rel}
        \begin{split}
            \hat{H}_{{\rm rel}, d}  = &- \frac{\hbar^2}{2\mu} \Delta_{r} + \frac{1}{2}\mu\omega^2(z^2+\eta^2\rho^2) - \mu\omega^2az + \frac{1}{2} \mu \omega^2 a^2 \\
            & - \frac{C_6}{r^6} - \frac{d^2(3\cos^2\theta - 1)}{4\pi\varepsilon_0 r^3}. \\
        \end{split}
    \end{equation}
    Note that we assume state-dependent $C_6$ and $d$ interaction parameters, and the presence of the trap does not mix these channels due to separation of energy scales between the rotational structure (GHz) and the tweezer frequencies (MHz).

    In the next step, we transform the above Hamiltonian into the set of coupled radial equations by expanding it using spherical harmonics with the following coupling rules
    \begin{equation*}
	    \bra{l,m}z\ket{l',m'} \sim \bra{l,m}Y^0_1\ket{l',m'} \sim \delta_{l',l \pm 1}
    \end{equation*}
	
    \begin{equation*}
	    \bra{l,m}V_{\rm dip}\ket{l',m'} \sim \bra{l,m}Y^0_2\ket{l',m'} \sim \delta_{l',l \pm 2}
    \end{equation*}

    Spherical harmonics provide an efficient basis when the trap separation is small or moderate, since the wavefunction has smooth angular structure and only a few partial waves contribute significantly. However, for separation $a \gg \xi_z$ larger than the characteristic trap length $\xi_z=\sqrt{\hbar/\mu\omega}$, the wavefunction acquires angular features of width $\Delta \theta \sim \xi_z/d$ which for proper description requires partial waves up to $l_{max} \sim a/\xi_z$. Hence, the number of relevant channels grows infinitely as $a \to \infty$. Fortunately, all physically relevant results are contained in the regimes of moderate displacement, while high displacement regime can be anticipated from analytical non-interacting solutions (cf. figure \ref{fig:trap_res_nodip}). 

    As a last step before solving the system of radial equations we need to discuss boundary conditions. We are interested in bound states so naturally at long range the wavefunction should vanish everywhere. On the contrary, due to the singular nature of the van der Waals interaction the short range boundary conditions are not that simple. We therefore start from the zero energy solution at a range where the van der Waals interaction is dominant ($C_6/r^6 \gg d^2/r^3, l(l+1)/r^2$)~\cite{Zdziennicki2025}. The solutions are expressed in terms of the Bessel functions 
    \begin{equation}\label{eq:asym_sol}
    	u_0(r) = A\sqrt{r} J_{\frac{1}{4}}\left(\frac{1}{2r^2}\right)
    	+ B\sqrt{r}Y_{\frac{1}{4}}\left(\frac{1}{2r^2}\right),
    \end{equation}
    which asymptotically behave as
    \begin{equation}
    	u_0(r) \sim \sin\left(\frac{1}{2r^2}-\frac{3}{8}\pi + \phi\right),
    \end{equation}
    where $\phi$ is a short-range phase that can be related to a physical parameter such as the $s$-wave scattering length $a_s$~\cite{Gribakin93,Gao98} in the spirit of Quantum Defect Theory. As the diagonal interaction dominates at short range, $\phi$ can effectively parametrize the behavior in all the channels.

	\subsection{Diagonalization procedure}\label{sec:diag}
\begin{figure}[H]
		\centering
		\includegraphics[width=\linewidth]{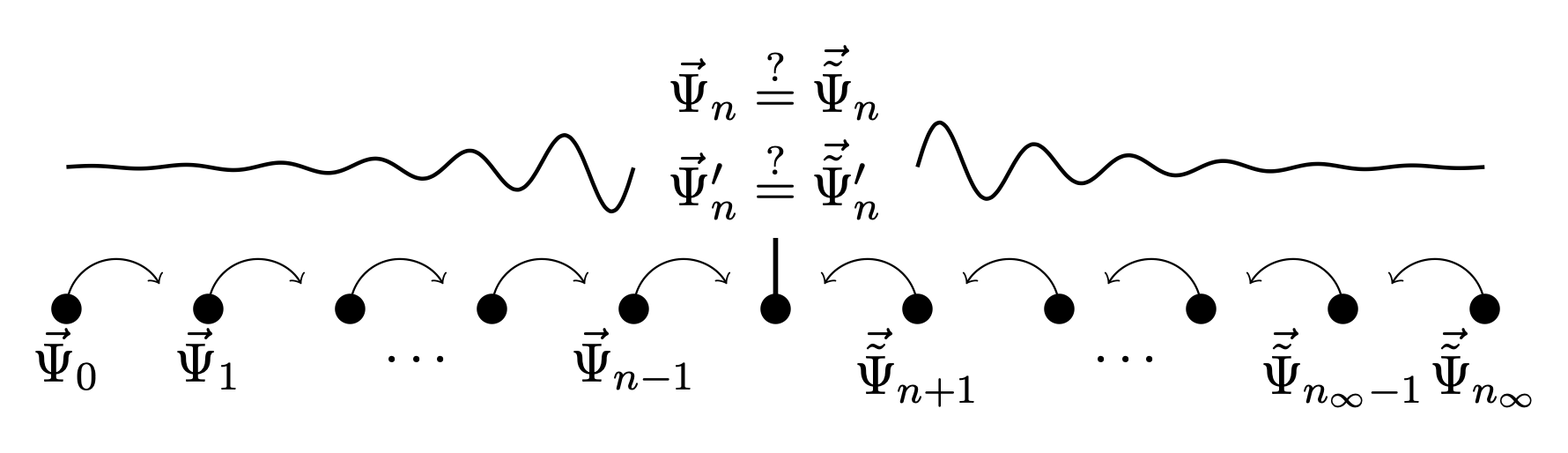}
		\caption{Pictorial representation of the Numerov algorithm. Starting from the solution at the boundary,
			we iteratively propagate the wavefunction from both sides to the intermediate matching point. If the initial energy used in propagation is close enough to the eigenvalue of the system, then continuity condition at the matching point can be met and this is the solution of the Schrödinger equation.
		}
		\label{fig:numerov_scheme}
	\end{figure}

	We then turn to the description of the external trapping potential. Here it is tempting to use harmonic oscillator basis which diagonalizes part of the Hamiltonian in a straightforward way~\cite{Sroczynska2021}. However, describing a tightly bound molecular state would then require very large basis size, as one needs to describe a narrow wavefunction with a set of Hermite polynomials with larger width given by the trap length scale, rendering the calculations inefficient and hard to converge in parts of the energy spectrum that we are the most interested in. Therefore, we numerically diagonalize Hamiltonian~\eqref{eq:H_rel} using renormalized Numerov method \cite{Johnson1978,DU1993}. The essence of the standard version of Numerov algorithm \cite{noumerov1924} consists of iterative propagation of the wavefunction from the boundary condition to the matching point (Fig. \ref{fig:numerov_scheme}). Iteration procedure can be carried out for an arbitrary initial energy, but continuity condition for the log-derivative at the matching point can be satisfied only for initial energy that is an eigenvalue of the Hamiltonian.
	
	The renormalized version proposed by Johnson extends the standard method to multichannel formalism, but also adds additional stability schemes: ratio matrix formulation, and node counting. The former reformulates the propagation scheme to the ratio of the wavefunction at the neighboring grid points, which is of the order of unity,	thereby increasing numerical stability especially in highly oscillatory regimes, while also simplifying the treatment of the log-derivative continuity condition. The latter provides a tool to keep track of the solutions using Sturm–Liouville theory. By counting negative diagonal elements of the ratio matrix,	we can assign a unique generalized principal number to every calculated state, ensuring that none of the solutions were missed. Johnson suggests that this procedure is proven to always work in the decoupled case, and gives correct results for the considered coupled multichannel cases. Here we supplement this scheme by diagonalization of the ratio matrix prior to counting its negative elements, this way reducing the problem locally to the decoupled case.
	
	Furthermore, we use an enhanced method of the matching point selection. According to Johnson’s condition, a proper matching point should be located far from a wavefunction node.  Due to the differing characteristics of eigenstates, it is difficult to define a single energy-independent matching point. Therefore, we choose the matching point at the location of the wide local extremum of the wavefunction, which should coincide with the point where the determinant of the ratio matrix is close to 1. In this way, the matching point can be determined on the fly during the iterative solution of the equations. By direct comparison of results from both of the schemes, we obtain further increase in stability of the results.

	The crucial step in calculations involving the singular van der Waals interaction is the proper implementation of the short-range boundary condition. The first task is to identify the region where the van der Waals interaction dominates. In our calculations, this region is defined as the range where the van der Waals potential exceeds other interaction terms by at least two orders of magnitude. For the systems considered here, the dipolar and van der Waals length scales overlap significantly. As a result, the onset of van der Waals dominance occurs at shorter distances than in the non-dipolar case. This has direct consequences for the numerical grid. Since the boundary condition is imposed through the wavefunction values at neighboring grid points, the grid spacing must be sufficiently small in the short-range region. If the spacing is too large, the potential may vary by orders of magnitude between adjacent points, effectively rendering it discontinuous at the numerical level and preventing stable integration.

	\section{Results}
	
	\subsection{Trap-induced resonances}
\begin{figure}[H]
		\centering
		\includegraphics[width=\linewidth]{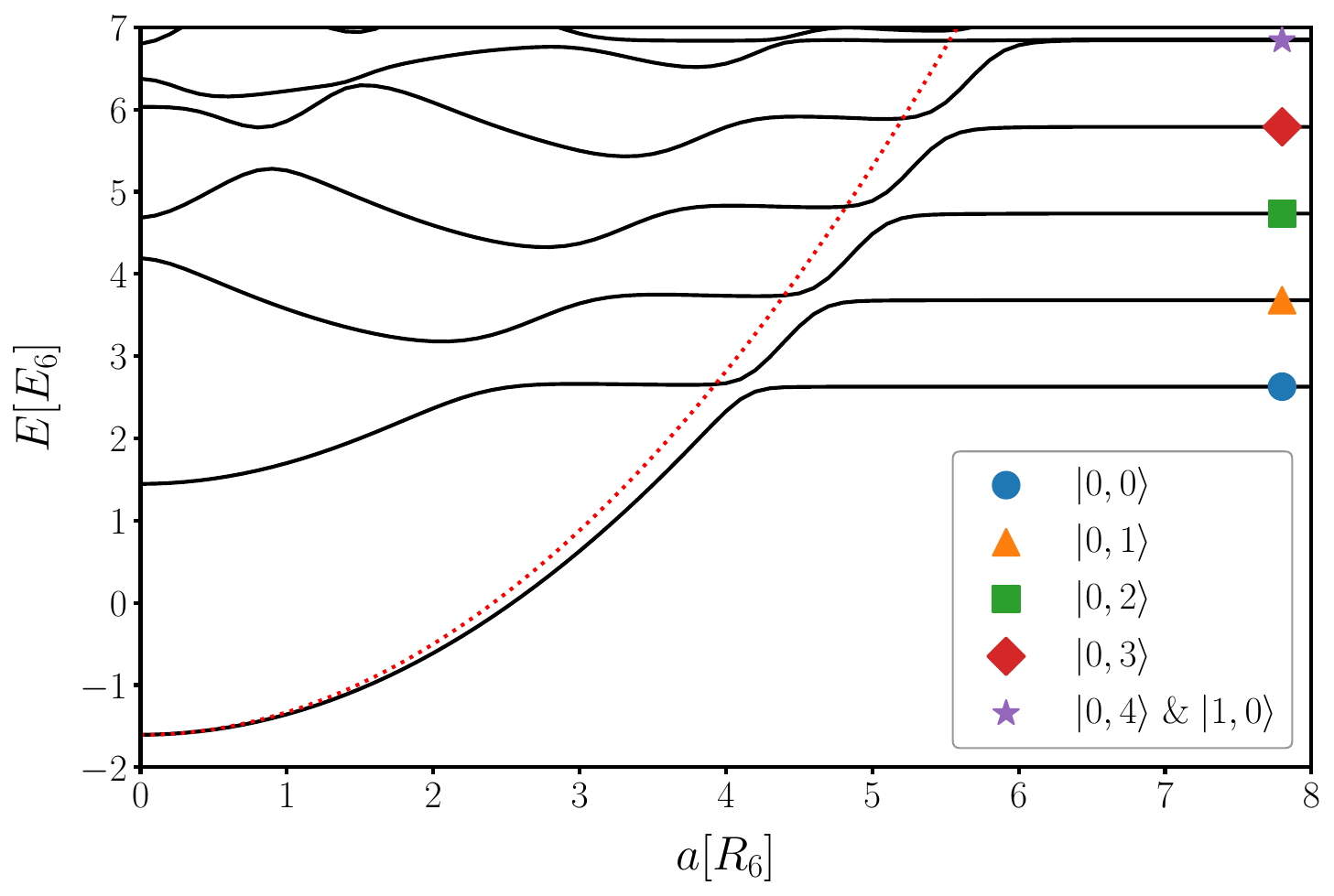}
		\caption{
			Energy levels as a function of tweezer separation for NaCs molecules with $\omega = 100$ kHz, $\eta = 2$,
			$C_6 = 8.21277 \cdot 10^6$ atomic units and $C_3 = 0$. The spectrum is produced for the angular momentum projection $m = 0$. The red dashed line denotes expected quadratic scaling of the molecular state energy, while the symbols mark the positions of unperturbed harmonic oscillator states.
			We denote them by $\ket{n_{xy}, n_z}$, where $n_{xy} = n_x = n_y$ which is a relation fullfilled for $m=0$.
			}
		\label{fig:trap_res_nodip}
	\end{figure}

	The trap-induced resonance occurs when for a specific trap configuration, the energy of a shallow molecular state equals to that of the trap vibrational state, causing them to hybridize and form an anti-crossing which can be wide if the overlap between the states is large. It results in a characteristic ripple in the energy level structure (see fig.~\ref{fig:trap_res_nodip}), where multiple states are being pushed upwards due to the level repulsion. The crossing occurs, as with increasing separation between the traps, the existence of a spatially localized molecular bound state becomes energetically expensive. Looking at the relative Hamiltonian (\ref{eq:H_rel}) one can expect to observe quadratic-like scaling of the molecular bound state energy
	\begin{equation}
		E_{{\rm mol},a} = \braket{\hat{H}_{{\rm rel},a}} = 
		E_{{\rm mol},0} - \mu \omega^2 a \braket{z} + \frac{1}{2}\mu \omega^2 a^2 , 
	\end{equation}
	where $\braket{z} \approx 0$ at short distances.

\begin{figure}[H]
		\centering
		\includegraphics[width=\linewidth]{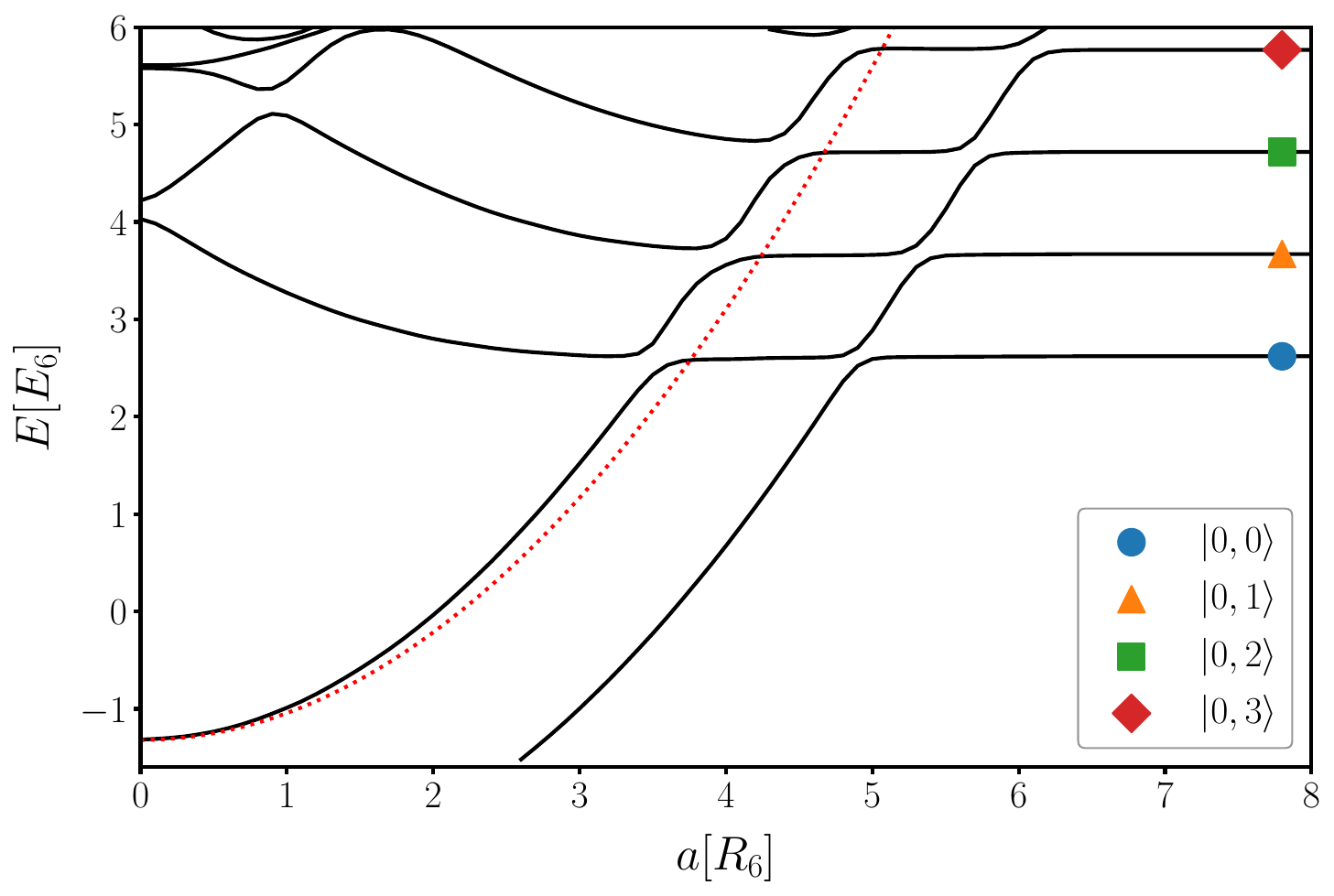}
		\caption{
			Energy spectrum for the same system as in Fig.~\ref{fig:trap_res_nodip}, but including moderate dipolar interaction ($R_3/R_6 = 1$).
			}
		\label{fig:trap_res_dip}
	\end{figure}

	If the distance between the traps is large and there are no bound states around, the interactions almost vanish, and the energy levels simply resemble the spectrum of two decoupled harmonic oscillators. Figure~\ref{fig:trap_res_nodip} shows a basic example where the trap-induced resonance occurs for just van der Waals interactions. Turning on even moderate dipole moment changes the nature of the interaction, which now supports a larger density of bound states due to the $\sim1/r^3$ tail. As a result, in Fig.~\ref{fig:trap_res_dip} we observe two trap-induced resonances. The other features of the energy spectrum remain qualitatively similar. 
	
	\subsection{Quantum gate protocol}
	The basic idea for implementing a quantum gate in the setup is to achieve state-dependent dynamics simply by moving the tweezers. When the trap separation becomes time-dependent, the dynamics can in principle be complex and require propagation of the full multichannel Hamiltonian. Quantum control methods can be used to handle the problem and achieve the desired gate even in the presence of many coupled channels by optimizing $a(t)$, and can include realistic assumptions such as experimentally achievable time resolution~\cite{Goerz2019,Muller2022}. Here, we restrict to a simple scheme in which the evolution is mostly adiabatic, and the crucial part of the process happens close to the trap-induced resonance position~\cite{Stock2006,Doerk2010}. Particle transport in tweezers can be arranged such that it is fast and no excitations in motional state are produced, again utilizing optimal control techniques~\cite{Cicali2025}. In order to illustrate the core of the gate operation, it is sufficient to use the Landau-Zener formula for the probability of diabatically passing the  resonance (which results in jumping between the curves in Figs.~\ref{fig:trap_res_nodip} and \ref{fig:trap_res_dip}) $P=\exp\left(-2\pi \Gamma/(|\dot{a}\partial \Delta E/\partial a|)\right)$, where $\Gamma=|\langle \psi_1|H|\psi_2\rangle|^2$ describes the overlap matrix element between the two states, $\dot{a}$ is the crossing speed, and $\partial \Delta E/\partial a$ the rate of change of the difference between the two energies. It is straightforward to control the outcome of the process by tuning the tweezer velocities. Importantly, we choose decoupled qubit states characterized by different angular momentum projections, which allows to study each state separately. The states differ only by the interaction coefficients $C_6$ and $C_3$, which are tunable with external fields.
	
	For demonstration, consider the energy spectra in Figs.~\ref{fig:trap_res_nodip} and \ref{fig:trap_res_dip}, which mainly differ by the positions of the TIRs. The difference is of the order of $R_6$, which is easily within experimental resolution. By bringing the tweezers to the vicinity of the first TIR slowly enough to follow the adiabatic energy curves, the two states acquire a difference in eigenenergies, which directly translates to accumulation of relative phase difference $\propto \int \Delta E(a(t))dt$, providing a route towards a controlled phase gate. The robustness criteria for this process are well known and have been optimized in earlier works~\cite{Doerk2010}.  The crucial aspect of the current protocol is that dipolar interactions are strong enough to matter on large length scales, making the problem numerically harder due to their anisotropy, but also offering larger couplings and thus wider anticrossings. Large tweezer frequency is also helpful for isolating excited states and allowing for speeding up the process: the adiabatic ``speed limit'' is set by the harmonic frequency as $\sqrt{\hbar\omega/\mu}$, so the achievable gate time for $\omega\sim$100kHz is in the submillisecond regime even without system-specific optimization. The state preparation and readout phases take place at large separation, where interactions can safely be neglected. Note that in principle deeply bound molecular levels cross with the trap energies at large distances, which may raise the question whether the molecules are ever truly separated. However, the overlaps for deeply lying states are negligibly small, such that the TIRs are extremely narrow and do not impact the system.
	
		\begin{figure}[H]
		\centering
		\includegraphics[width=\linewidth]{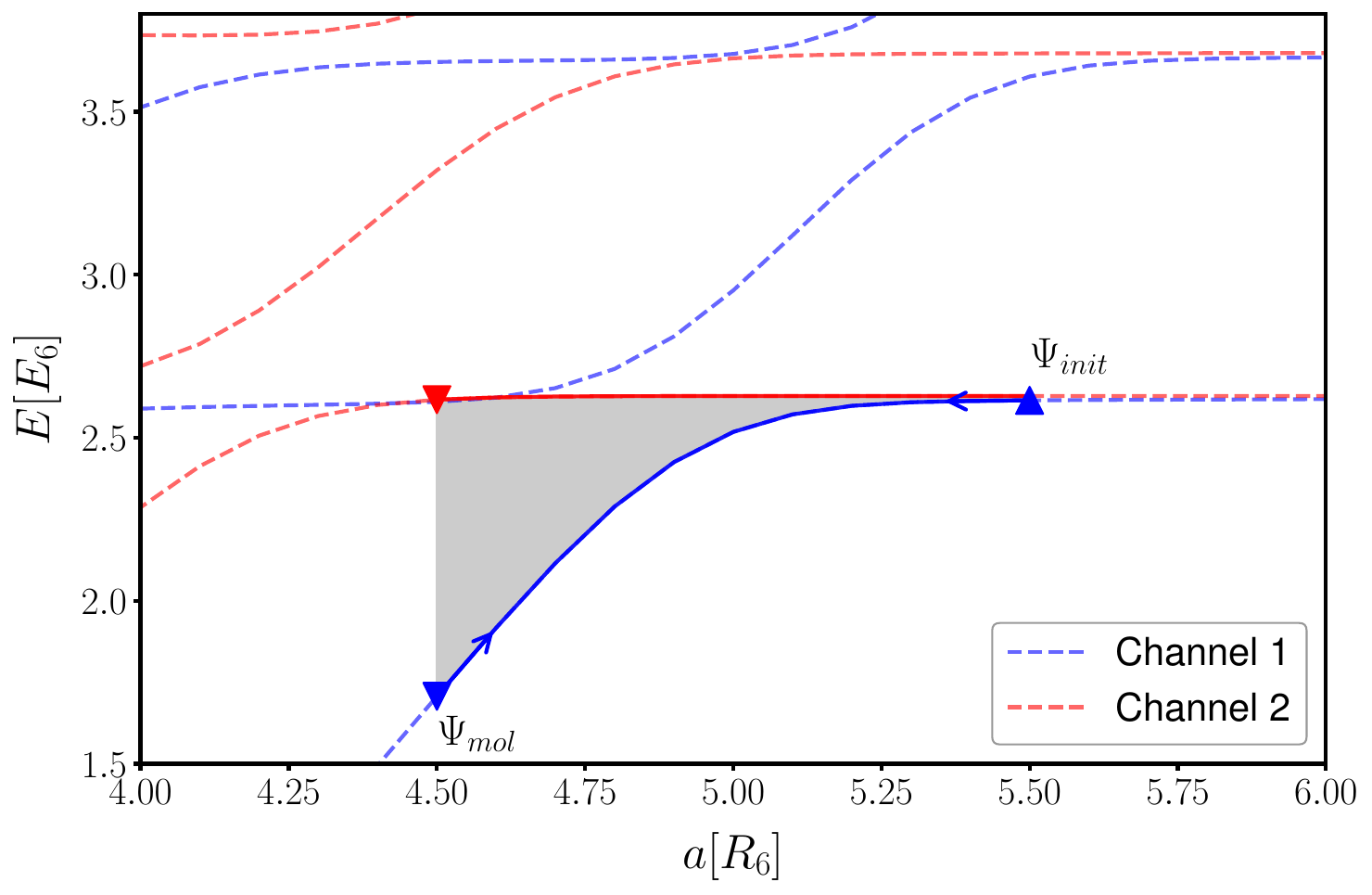}
		\caption{
			The basic mechanism behind the quantum gate protocol, depicting the dependence of the energy difference on the tweezer separation for the two channels.
		}
		\label{fig:gate}
	\end{figure}

	\subsection{Resonantly enhanced quantum sensing} 
	
	Trap-induced resonances could also be leveraged for detection of external fields. Here, we envision two simple protocols which are only based on moving the traps. In the first one, initially well separated molecules are abruptly brought to a short relative distance, and then slowly separated again. If a trap-induced resonance is present, the particles will cross it diabatically on one way, and adiabatically on the way back, leading to a detectable excitation of the relative motion. This approach allows for detecting whether the field exceeds a critical value at which the resonance appears at certain separation.
	
	In the second protocol, the traps would be brought to a distance $a_{\rm min}$ and then back, but this time sufficiently slowly to be adiabatic in both directions. The accumulated phase during this process depends on the presence of an avoided crossing. This phase can be detected using an interferometer with a different molecular state serving as a phase reference. The range of detectable fields can be tuned by changing the smallest separation during the process, as well as by the external trap parameters. The precision would naturally increase with the time spent at shortest separation where the difference in energies between the two states is the largest. The sensitivity of such a sensor relies on the variation of the TIR position with the field strength. Here, the fluctuations in the tweezer positions would be the primary source of uncertainty.

	\section{Conclusions}

	This work numerically resolves the two-body energy structure of interacting ultracold polar molecules confined in optical tweezers, capturing the interplay between short and long-range interactions and the external trapping potential. Using a coupled-channel approach based on the renormalized Numerov method, we show that trap-induced resonances arise naturally for experimentally relevant trap separations and interaction strengths. Moreover, their positions and widths can be predicted with sufficient accuracy to allow these resonances to be used as a resource for quantum control rather than a limitation.
	
	These results indicate that the confining structure of optical tweezers can serve as an active ingredient for quantum state engineering of molecules. By mapping the interaction landscape onto an effective two-qubit description, we outlined an adiabatic controlled-phase gate based on state-dependent trap-induced resonances. Although the basic mechanism parallels earlier trap-induced gate schemes developed in other platforms, the molecular case offers favorable conditions because dipolar interactions remain significant on comparatively large length scales and generate broader avoided crossings, thereby expanding the useful operating window.
	
	Beyond quantum gate applications, the same resonance structure offers a natural pathway to resonantly enhanced sensing. Because the positions of trap-induced resonances are sensitive to external fields, both motional excitation and phase-sensitive measurements can serve as probes for weak electric field variations. More broadly, these results identify trap-induced resonances as a versatile tool for quantum control in molecular tweezer platforms. Future research should include full dynamical simulations of the gate process, optimization in the presence of realistic imperfections, and extension of the present two-body framework to larger molecular arrays for programmable quantum information processing and sensing.

	\emph{Acknowledgements.} This work was supported by the National Science Centre of Poland Grant 2020/37/B/ST2/00486.
	
	\bibliography{mol_gate_bib}
	
\end{document}